\documentclass[journal]{IEEEtran}
\usepackage{graphicx}
\usepackage{amsmath}
\usepackage{color}
\usepackage{nopageno}
\usepackage{multicol}
\usepackage{wrapfig}

\begin{document}
\pagenumbering{gobble}
\title{Formation of Solitary Structures and Double Layer in Weakly Degenerate Electron-Ion Quantum Plasma in Ionosphere}

\author{J Sarkar, J Goswami, S Chandra and B Ghosh
\thanks{\hrule{}\vspace{1mm} Jit Sarkar is with Department of Physics, Jadavpur University, Kolkata - 700 032, India. (e-mail: sarkarjit101@gmail.com)}
\thanks{Jyotirmoy goswami is with Department of Physics, Jadavpur University, Kolkata - 700 032, India. (e-mail: jyotirmoygoswami09@gmail.com)}
\thanks{Swarniv Chandra is with Govt. General Degree College at Kushmandi, Dakshin Dinajpur, 733121, India.(e-mail: swarniv147@gmail.com)}
\thanks{Basudev Ghosh is with Department of Physics, Jadavpur University, Kolkata - 700 032, India.(e-mail: bsdvghosh@gmail.com)}}

\maketitle
\begin{abstract}
In this paper, we consider ionospheric plasma consisting of weakly degenerate electrons and heavy ions. We embrace our hydrodynamic model by including the quantum diffraction term. By employing Sagdeev's pseudo-potential method, we obtain double layers and soliton structure. We have studied the various parametric dependence of solitary structures and double layers. The results thus obtained might be helpful in the studies of many high energy astrophysical phenomena.
\end{abstract}

\section{Introduction}
\IEEEPARstart{D}{\lowercase{ouble layers}} (DLs) are structures in a plasma which consists of two oppositely charged sheet that has a substantial potential difference between them. They thus have a relatively strong field in between them compared to the global electric fields. Particles can thus be accelerated, decelerated, or drifted as par their initial direction of motion. DLs are created in laboratory-produced plasmas \cite{levine1980fluid} as well as found in space plasmas \cite{block1978double, hasan1978alfven, knorr1974existence,akbari2016double}. The magnetospheric plasma shocks are an offshoot of DLs, and the aurora is regions where DLs are obscured. Much like the shocks, DLs can be classified as weak and strong based on their strength. Depending upon their energy drops, which are comparable to particles' rest mass energy, they are either relativistic or non-relativistic \cite{carlqvist1982physics, andrews1971theory,albert1970auroral}. Based on factors like current conduction, there are current-carrying DL's displaying Ferlay-Bunemann instability \cite{schlegel1994observation, dimant1995kinetic}.
\par
Regarding the morphology of DL's, there are four regions-
\begin{enumerate}[(a)]\itemsep-1pt
\item A positive potential side of DL's towards which electrons are accelerated.
\item A positive potential where electrons are retarded.
\item A negative potential side where electrons are decelerated.
\item a negative potential side where electrons are accelerated.
\end{enumerate}

\par
Magnetospheric DL is a transient feature as there is a necessity of constant energy to sustain such structure. Dls are formed by a variety of mechanisms like injecting electron current \cite{wescott1976skylab}, increasing current density, instability, parity effects, etc.
\par
This chapter will discuss the foundation of DL and solitary structure formations in an electron-ion plasma where electrons follow Fermi distribution. We have obtained a Sagdeev's pseudo-potential feature by non-perturbative technique, and after that, we used proper boundary conditions to obtain the solitary structure and double layer. The findings of the work will be helpful in many laboratory plasmas and space plasmas phenomena. In section III, we derive the expression for Sagdeev's pseudo-potential. In the following section, we discuss the structure.


\section{Basic equations}
We consider a homogenous, un-magnetised, two component quantum plasma consisting inertia-less electron, and ion. The  basic  equations  describing the nonlinear dynamics of the ion acoustic wave in the quantum plasma system are
\begin{equation}\label{eq121}
\frac{\partial {{n}_{e}}}{\partial t}+\frac{\partial \left( {{n}_{e}}{{v}_{e}} \right)}{\partial x}=0
\end{equation}
\begin{equation}\label{eq122}
\frac{\partial {{n}_{i}}}{\partial t}+\frac{\partial \left( {{n}_{i}}{{v}_{i}} \right)}{\partial x}=0
\end{equation}
\begin{equation}\label{eq123}
0=\frac{e}{{{m}_{e}}}\frac{\partial \phi }{\partial x}-\frac{1}{{{m}_{e}}{{n}_{e}}}\frac{\partial {{p}_{e}}}{\partial x}+\frac{{{h}^{2}}}{2m_{e}^{2}}\frac{\partial }{\partial x}\left[ \frac{{{\partial }^{2}}\sqrt{{{n}_{e}}}/\partial {{x}^{2}}}{\sqrt{{{n}_{e}}}} \right]
\end{equation}
\begin{equation}\label{eq124}
\frac{\partial {{v}_{i}}}{\partial t}+{{v}_{i}}\frac{\partial {{v}_{i}}}{\partial x}=-\frac{e}{{{m}_{i}}}\frac{\partial \phi }{\partial x}-\frac{1}{{{n}_{i}}}\frac{\partial {{p}_{i}}}{\partial x}
\end{equation}
\par And above mentioned system bounded by Poison's equation, which in this case is
\begin{equation}\label{eq125}
\frac{{{\partial }^{2}}\phi }{\partial {{x}^{2}}}=-4\pi e\left( {{n}_{i}}-{{n}_{e}} \right)
\end{equation}
where, $n_j$, $v_j$, $m_j$ and $e$ are the density, velocity, mass of $j-th$ species ($j=e,i$) and charge respectively. $\hbar$ is the Planck constant, $\phi$ electro static potential, $p_j$ is respective species pressure effect. And we assumed  ${{p}_{e}}=\frac{1}{8}{{\left( \frac{3}{\pi } \right)}^{\frac{1}{3}}}\hbar cn_{e}^{\frac{4}{3}}$ and ${{p}_{i}}=n{{k}_{B}}T$ where $k_b$ is Boltzmann constant and $T$ is temperature.
\par For convenience, we introduce the following normalisation scheme $x\to \frac{x{{\omega }_{i}}}{{{c}_{s}}}$, $t\to t{{\omega }_{i}}$, $\phi \to \frac{e\phi }{2{{k}_{B}}{{T}_{Fe}}}$, ${{n}_{j}}\to \frac{{{n}_{j}}}{{{n}_{0}}}$, ${{u}_{j}}\to \frac{{{u}_{j}}}{{{c}_{s}}}$ where, ${{\omega }_{e}}=\sqrt{4\pi {{n}_{e0}}{{e}^{2}}/{{m}_{e}}}$ and ${{c}_{s}}=\sqrt{2{{k}_{B}}{{T}_{Fe}}/{{m}_{e}}}$.
\par The normalized electron and ion momentum equation and ion continuity equation can be written as
\begin{equation}\label{eq126}
\frac{\partial {{n}_{e}}}{\partial t}+\frac{\partial \left( {{n}_{e}}{{v}_{e}} \right)}{\partial x}=0
\end{equation}
\begin{equation}\label{eq127}
\frac{\partial {{n}_{i}}}{\partial t}+\frac{\partial \left( {{n}_{i}}{{v}_{i}} \right)}{\partial x}=0
\end{equation}
\begin{equation}\label{eq128}
0=\frac{\partial \phi }{\partial x}-\frac{4}{3}{{R}_{e}}n_{e}^{\frac{-2}{3}}\frac{\partial {{n}_{e}}}{\partial x}+\frac{{{H}^{2}}}{2}\frac{\partial }{\partial x}\left[ \frac{{{\partial }^{2}}\sqrt{{{n}_{e}}}/\partial {{x}^{2}}}{\sqrt{{{n}_{e}}}} \right]
\end{equation}
\begin{equation}\label{eq129}
\frac{\partial {{v}_{i}}}{\partial t}+{{v}_{i}}\frac{\partial {{v}_{i}}}{\partial x}=-\mu \frac{\partial \phi }{\partial x}-{\sigma}{{n}_{i}}^{-1}\frac{\partial {{n}_{i}}}{\partial x}
\end{equation}
\begin{equation}\label{eq1210}
\frac{{{\partial }^{2}}\phi }{\partial {{x}^{2}}}=\left( {{n}_{i}}-{{n}_{e}} \right)
\end{equation}
where, ${{R}_{e}}=\frac{1}{8}{{\left( \frac{3}{\pi } \right)}^{\frac{1}{3}}}hc$, $\sigma=3(T_{Fi}/T_{fe})$, $\mu=n_{e0}/n_{i0}$ and $H$ is non-dimensional quantum diffraction parameter.
\section{Nonlinear analysis}
In order to investigate localized stationary solution, we transform all variables into one and independent variable $\xi=x - Mt$ where M is Mach number.
Integrating eqs. \ref{eq128} and using proper boundary condition, we yield
\begin{equation}\label{eq1211}
\phi=-4{{R}_{e}}+4{{R}_{e}}n_{e}^{\frac{1}{3}}-\frac{{{H}^{2}}}{2}\frac{1}{\sqrt{{{n}_{e}}}}\frac{{{\partial }^{2}}\sqrt{{{n}_{e}}}}{\partial {{x}^{2}}}
\end{equation}
From eq. \ref{eq127} \& \ref{eq129} with boundary conditions ${{n}_{i}}\to 1$, $v_i\to 0$ and $\phi \to 0$ at $\xi =|\pm \infty |$, we obtain the following equations,
\begin{equation}\label{eq1212}
{{v}_{i}}=M\left( 1-\frac{1}{{{n}_{i}}} \right)
\end{equation}
and
\begin{eqnarray}\label{eq1213}
{{M}^{2}}{{\left(1-\frac{1}{{{n}_{i}}}\right)}^{2}}-2{{M}^{2}}\left(1-\frac{1}{{{n}_{i}}}\right)+{{\sigma }_{1}}\ln ({{n}_{i}})&=&-2\mu \phi\nonumber \\
 \Rightarrow \frac{1}{\mu }\left[ \frac{{{M}^{2}}}{2}\left( 1-\frac{1}{n_{i}^{2}} \right)-\frac{{{\sigma }_{1}}}{2}\ln ({{n}_{i}}) \right]&=&\phi
\end{eqnarray}
Now, employing quasi-neutrality condition $(n_i\approx n_e=n)$ and equating eq. \ref{eq1212} \& eq. \ref{eq1213}, we get
\begin{multline}\label{eq1214}
-4{{R}_{e}}+4{{R}_{e}}n_{e}^{\frac{1}{3}}-\frac{{{H}^{2}}}{2}\frac{1}{\sqrt{{{n}_{e}}}}\frac{{{\partial }^{2}}\sqrt{{{n}_{e}}}}{\partial {{x}^{2}}}=\frac{{{M}^{2}}}{2\mu }\left( 1-\frac{1}{n_{i}^{2}} \right)\\-\frac{{\sigma}}{2\mu }\ln ({{n}_{i}})
\end{multline}
Substituting $z=n^{1/2}$ and perform integration with appropriate boundary conditions i.e; $\ddot{n(\xi)}\to 0$, $\dot{n{(\xi)}}\to 0$ and $n\to 0$ as $\xi\to|\pm \infty|$  on eq. \ref{eq1213}, one obtain the energy integral equation as
\begin{equation}\label{eq1215}
\frac{1}{2}{{\left( \frac{dn}{d\xi } \right)}^{2}}+U(n)=0.
\end{equation}
where $U(n)$ can be expressed as
\begin{multline}\label{eq1216}
U(n)=\\\frac{2}{{{H}^{2}}}\left[ 2\left( {{R}_{e}}-\frac{{{M}^{2}}}{4\mu }-\frac{{\sigma }}{4} \right){{n}^{2}}-\frac{3}{2}{{R}_{e}}{{n}^{7/3}}+\frac{{{M}^{2}}}{4\mu }n\right]+\\\frac{2}{{{H}^{2}}}\left[\frac{{\sigma }}{4\mu }\left\{ n-{{n}^{2}}\ln (n) \right\}-\frac{{{R}_{e}}+\left( {{{M}^{2}}}/{\mu } \right)}{2}n \right]
\end{multline}
\par This equation describes the Sagdeev's pseudo-potential well $U(n)$ by which a particle's motion executed as a function of $(n)$. To yield soliton solution, we use some boundary conditions as follows:
\begin{multline}\label{eq1217}
a)\quad U(n)=0\quad at\quad  n=0\quad  and\quad  n=n_{m}\\
b)\quad \frac{d U}{d n}=0 \quad at\quad  n=0 \quad but\quad  \frac{d U}{d n} \neq 0 \quad at\quad  n=n_{m}\\
c) \quad \frac{d^{2} U}{d n^{2}}<0 \quad at\quad  n=0\qquad\qquad\qquad\qquad\quad
\end{multline}
In this case, the condition will be ($\sigma<1-M^2$) which implies that in this electron-ion quantum plasma only subsonic solitary wave can be formed. Structure of solitary wave depends on the value of $n_m$.
\par From eq. \ref{eq1215}, we can say that shape of soliton structure can be decided from the following equation:

\begin{equation}\label{eq1218}
  \xi=\pm\int_{n_m}^{n}\frac{dn_x}{\sqrt{-2U(n)}}
\end{equation}
\section{Solitary-wave solution}
To derive soliton solution, we expand $U(n)$, and assuming that $(n)$ is very small. So we considering smallest two order of $(n)$.
\begin{equation}\label{eq1219}
\frac{d^2 \phi}{d \xi^2}=A\phi+B\phi^2
\end{equation}
where, $A=\frac{\sigma}{2H^2\mu}$ and $B=\big(\frac{5\sigma}{12H^2\mu}+\frac{4R_e}{H^2}-\frac{M^2}{\mu H^2}\big)$.
\par Using transformation like $\phi(\xi)=z(w)$ and $w=sech\zeta(\xi)$, equation \ref{eq1219} reduces to

\begin{equation}\label{eq1220}
\zeta^2w^2(1-w^2)\frac{d^2z}{dw^2}+\zeta w(1-2w^2)\frac{dz}{dw}-Az-Bz^2=0
\end{equation}

As we found a regular singularity at $w=0$, we try to find the solution using power series solution in form of
\begin{equation}\label{eq1221}
z(w)=\sum_{r=0}^{\infty}b_rw^r
\end{equation}
The power series truncates at $r=3$, if $\zeta=\sqrt{A/4}$, $b_2=3A/2B$, $b_0=0$ and $b_1=0$, we get the soliton solution in form,
\begin{equation}\label{eq1222}
\phi(\xi)=\frac{3A}{2B}sech^2(\xi/\delta)
\end{equation}
where, $\delta=\sqrt{4/A}$. $\frac{3A}{2B}$ determines amplitude of solitay wave, and width of soliton depends on $\delta$.
\par Now we consider $\phi^{7/3}$ in $\frac{d^2\phi}{d\xi^2}$, we get
\begin{equation}\label{eq1223}
\frac{d^2 \phi}{d \xi^2}=A\phi+B\phi^2+C\phi^{7/3}
\end{equation}
where, $C= -\frac{3R_e}{H^2}$.
\par The boundary conditions for the formation of double layer are,
$$\bigg(\frac{d\phi}{d \xi}\bigg)^2=0\quad\text{at}\quad\phi=0\quad\text{and}\quad\phi=\phi_m$$
\par Applying the hyperbolic method($tanh$), we yield the double layer solution as,
\begin{equation}\label{eq1224}
\phi(\xi)=\phi_m[1\pm tanh(\xi/\Delta)]
\end{equation}
where, $\delta=3\sqrt{4C}/B$.
\section{Results and discussion:}

\begin{figure}
\centering
  \includegraphics[width=3.8in,height=80mm]{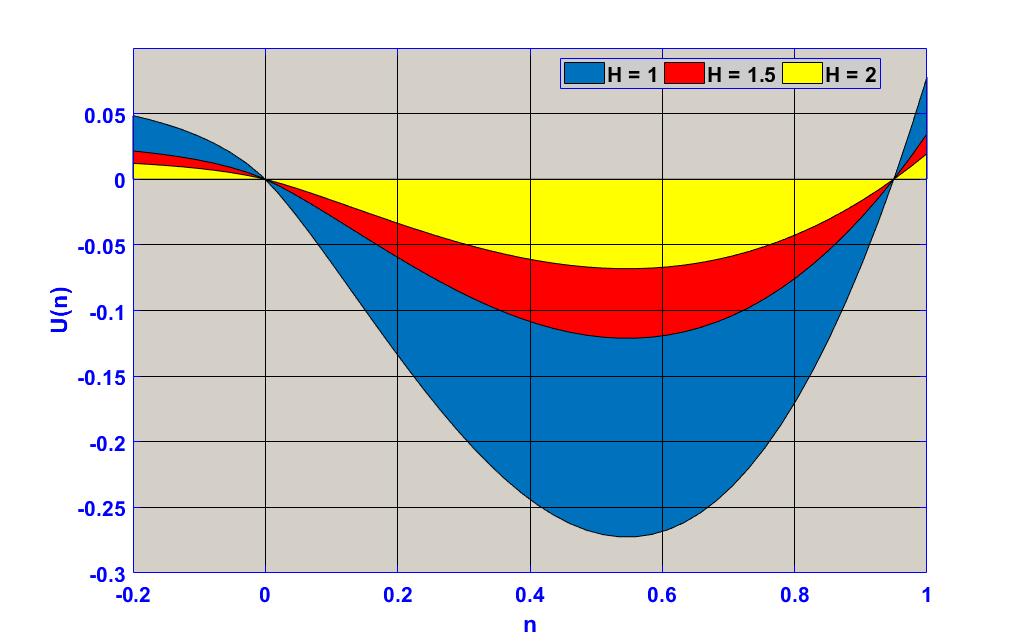}
  \caption{Plot of $U(n)$ for different values of diffraction parameter $H$ for M = 1.4, $R_e$=0.5 and $\mu$=0.6}\label{potentialH.jpg}
\end{figure}

\begin{figure}
  \centering
  \includegraphics[width=3.8in,height=80mm]{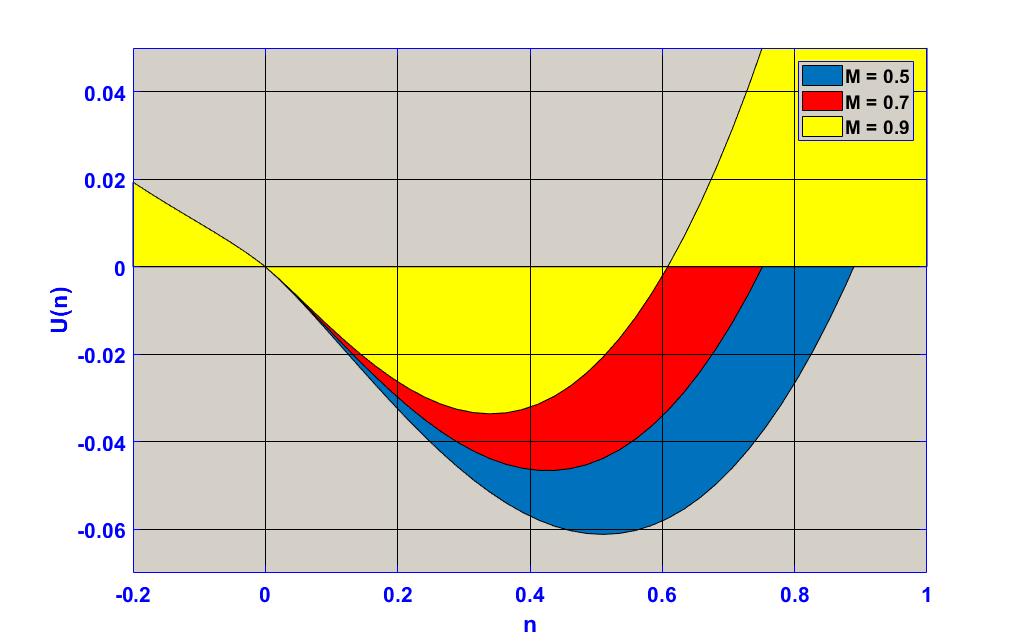}
  \caption{Plot of $U(n)$ for different values of Mach number $M$ for H = 2, $R_e$=0.5 and $\mu$=0.6}\label{potentialM.jpg}
\end{figure}

\begin{figure}
  \centering
  \includegraphics[width=3.8in,height=80mm]{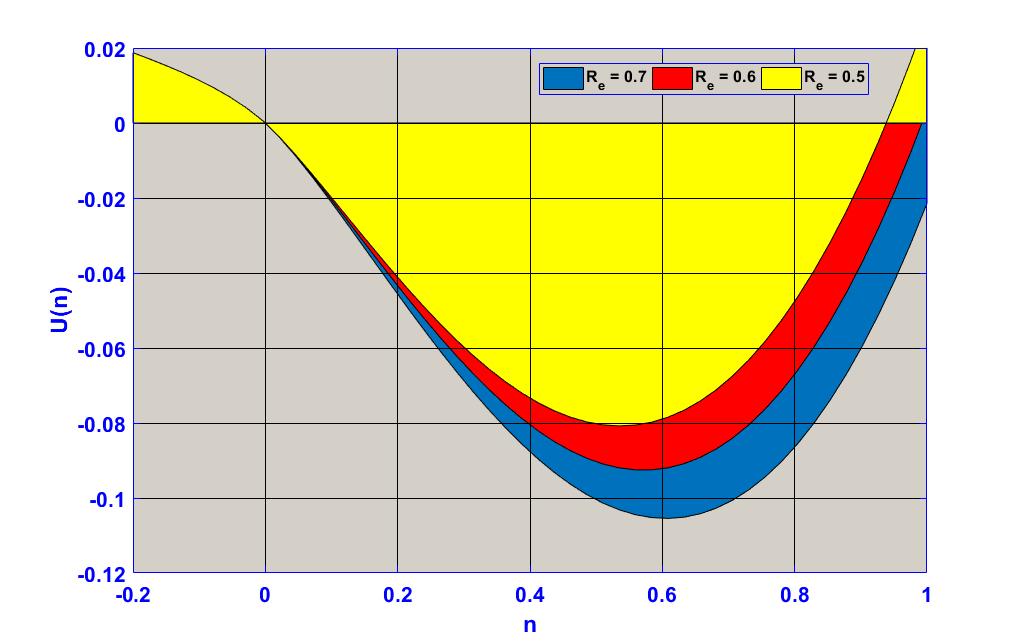}
  \caption{Plot of $U(n)$ for different values of $R_e$ for M = 1.4, $H$=2 and $\mu$=0.6}\label{potentialRe.jpg}
\end{figure}

\begin{figure}
  \centering
  \includegraphics[width=3.8in,height=80mm]{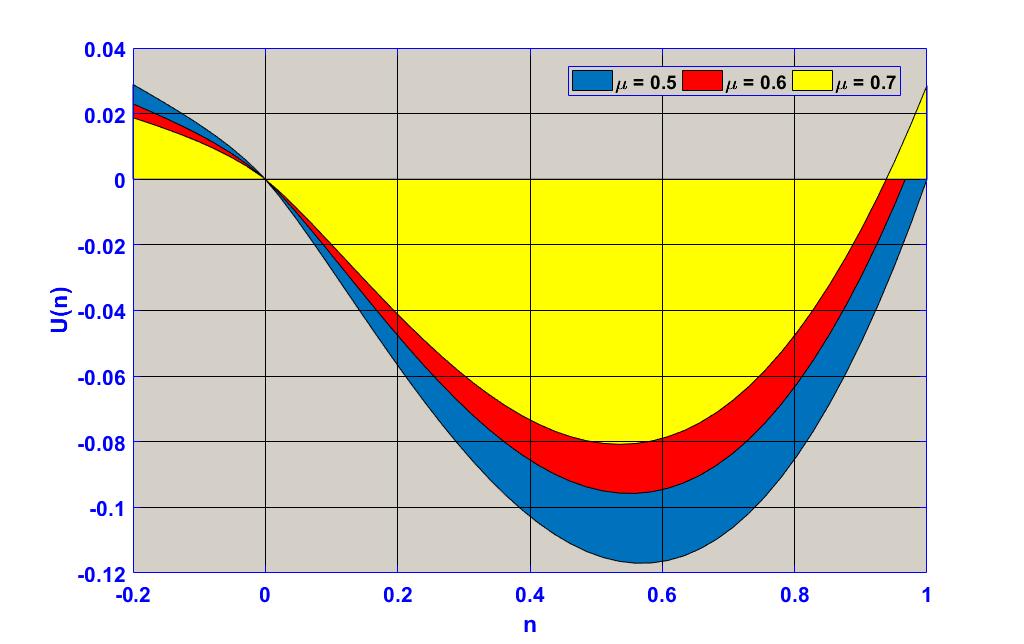}
  \caption{Plot of $U(n)$ for different values of $\mu$ for M = 1.4, $R_e$=0.5 and $H$=2}\label{potentialmu.jpg}
\end{figure}

\begin{figure}
  \centering
  \includegraphics[width=3.6in,height=80mm]{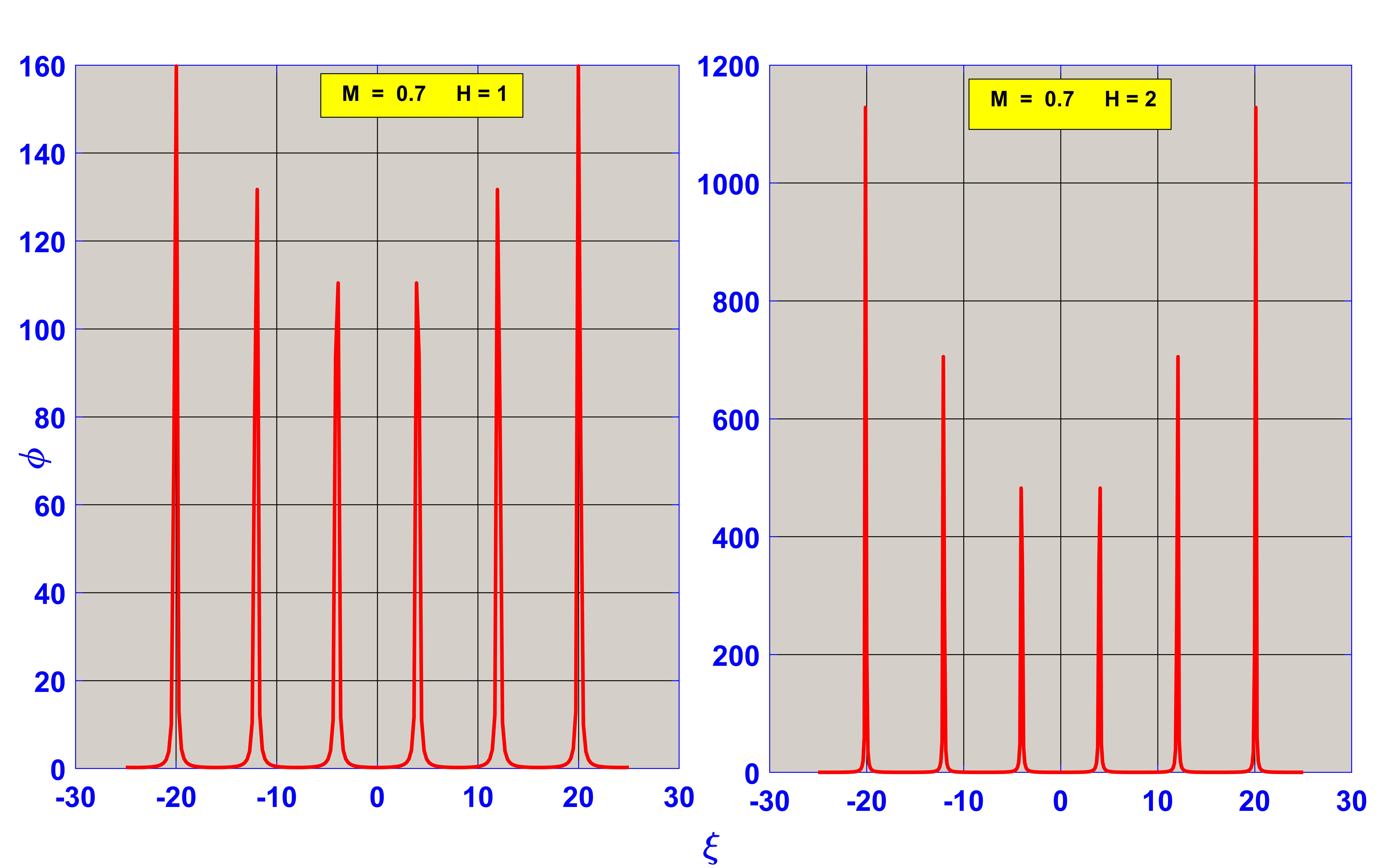}
  \caption{Fluctuations in the wave potential are depicted for different diffraction parameters H for $R_e$=0.5 and $\mu$=0.6}\label{solitaryH.jpg}
\end{figure}

To investigate the nature and formation of double layers and solitary structure, we plot the pseudo-potential function with the variation of different parameters and study its influence. Later we obtain the corresponding solution for double layer and solitary structures and plot for different parameters. In figure \ref{potentialH.jpg}, we find that the pseudo-potential function becomes shallow for increasing the value of the quantum diffraction effect. It implies that the higher the value of H, the more will be the electron tunneling effect, and the particles will be unable to form stationary structures. Figure \ref{potentialM.jpg} shows that the higher the wave speed (M) for a small range of particle (ion) density, there will be a static function. As the wave in plasma propagating gets little time for few particles to interact nonlinearity, it forms a stationary structure. In figure \ref{potentialRe.jpg}, the relativistic degeneracy parameter ($R_e$) is shown to give more nonlinear effects. For greater values of $'R_e'$, we get a more fantastic range of particle density over which stationary formation is possible. The mass ratio of electron to ion ($\mu$) has no such effect (Fig. \ref{potentialmu.jpg}) in delimiting the particle density necessary for obtaining stationary formations. The only thing is that the degree of nonlinearity is reduced slightly. Now we have further shown that the fluctuations are more pronounced in the quantum region by orders of tens (Fig. \ref{solitaryH.jpg})

\begin{figure}
  \centering
  \includegraphics[width=3.8in,height=80mm]{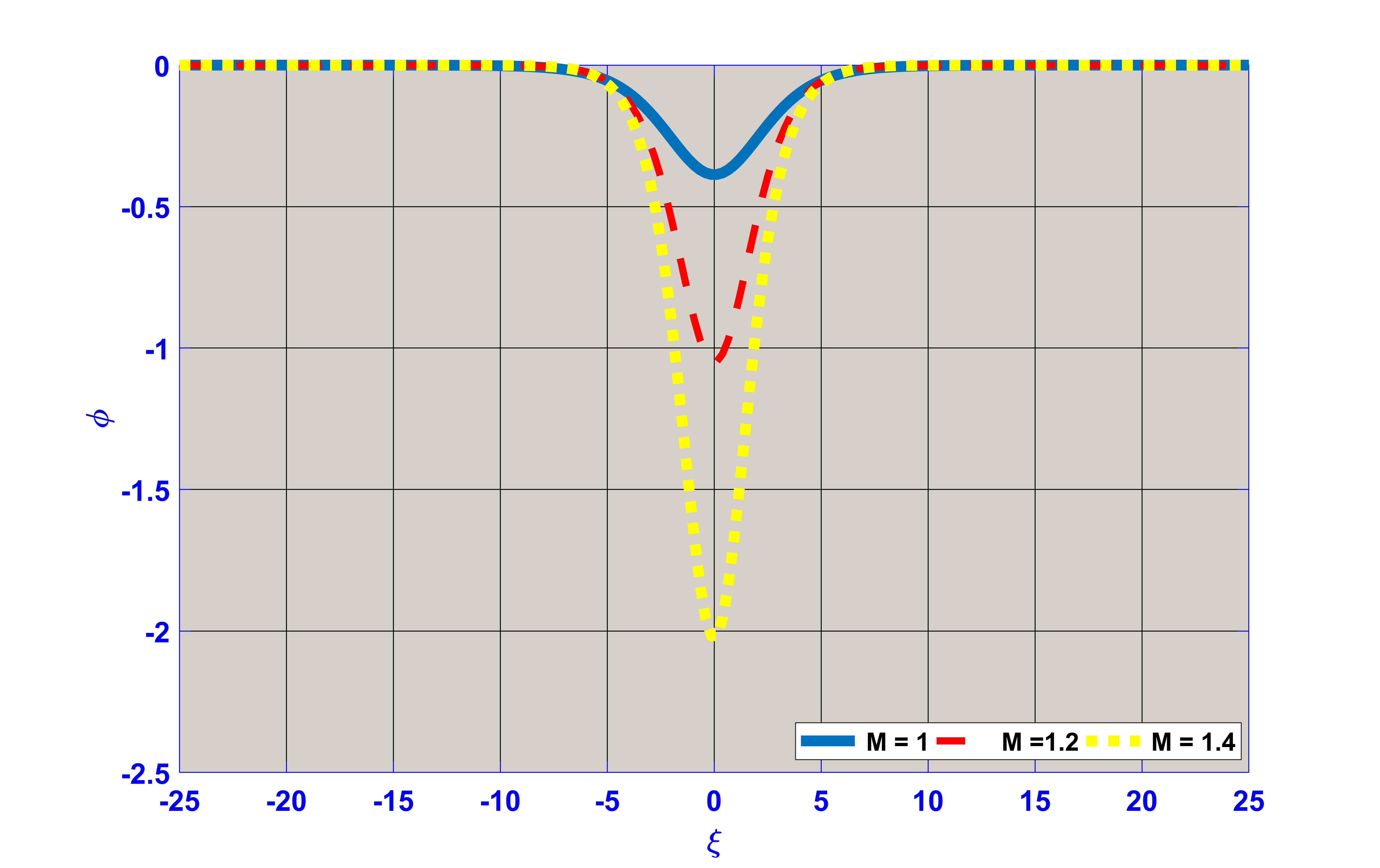}
  \caption{Soliton plot in $(\phi-\xi)$ space for different for different Mach number $M$ for H = 2, $R_e$=0.5 and $\mu$=0.6}\label{solitaryM.jpg}
\end{figure}

\begin{figure}
  \centering
  \includegraphics[width=3.8in,height=80mm]{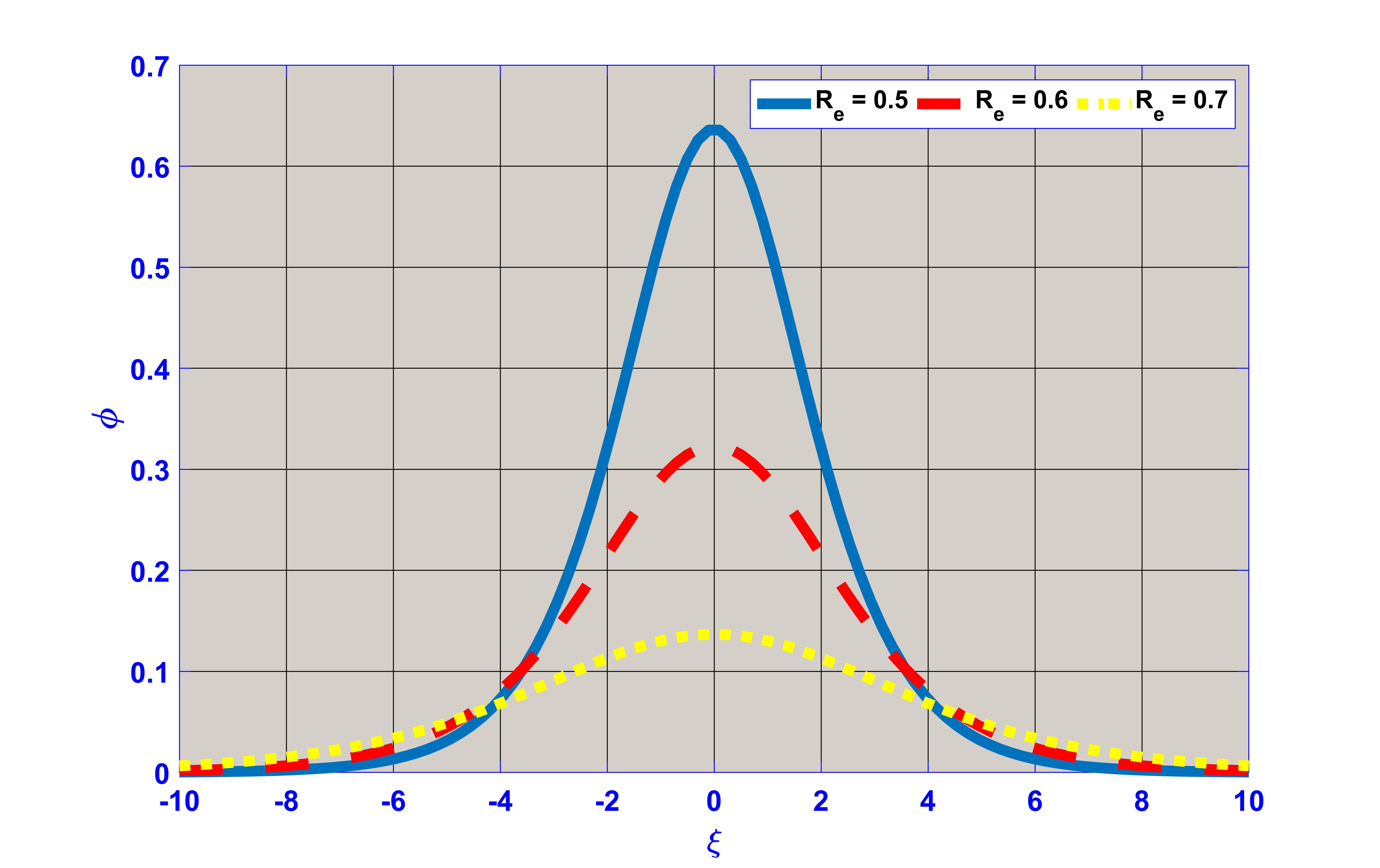}
  \caption{Soliton plot in $(\phi-\xi)$ space for different for different Mach number $R_e$ for M = 1.4, $R_e$=0.5 and $\mu$=0.6}\label{solitaryRe.jpg}
\end{figure}
\par
\begin{figure}
  \centering
  \includegraphics[width=3.8in,height=80mm]{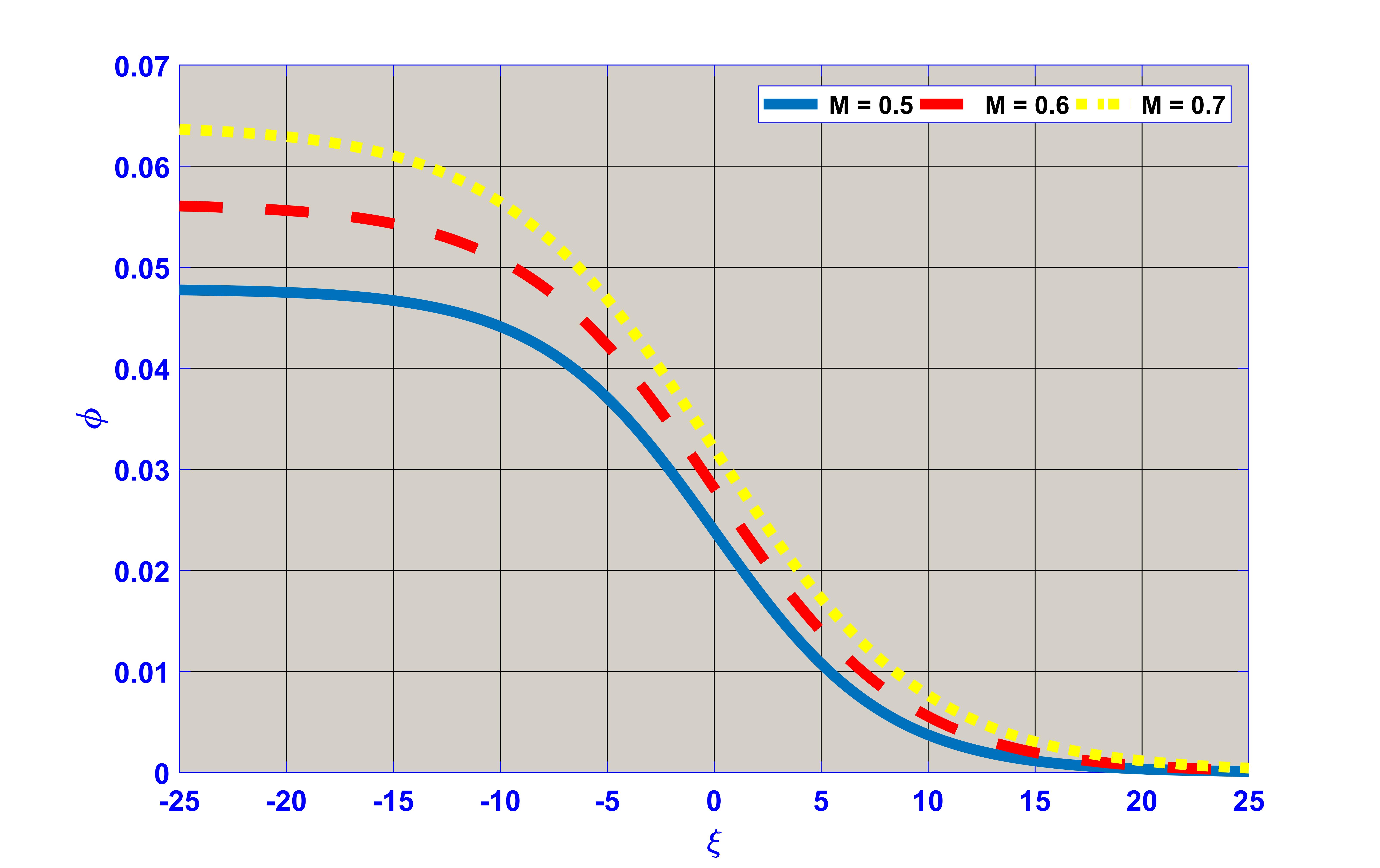}
  \caption{Double layer structure in $(\phi-\xi)$ space for different Mach number $M$ when H = 2, $R_e$=0.5 and $\mu$=0.6}\label{shockM.jpg}
\end{figure}
\begin{figure}
  \centering
  \includegraphics[width=3.8in,height=80mm]{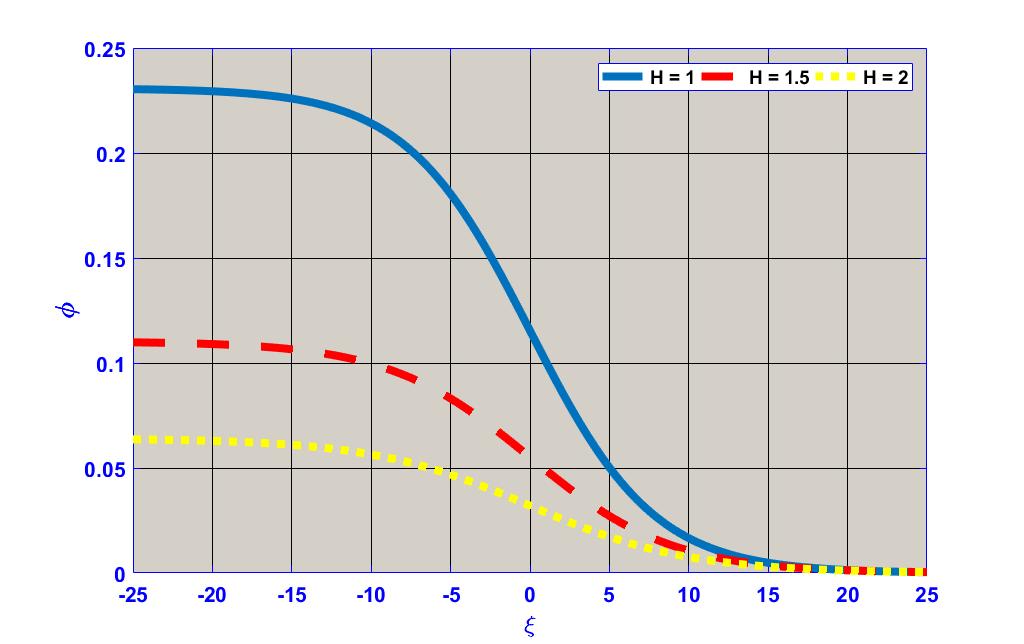}
  \caption{Double layer structure in $(\phi-\xi)$ space for different quantum parameter $H$ when M = 1.4, $R_e$=0.5 and $\mu$=0.6}\label{shockH.jpg}
\end{figure}
\begin{figure}
  \centering
  \includegraphics[width=3.8in,height=80mm]{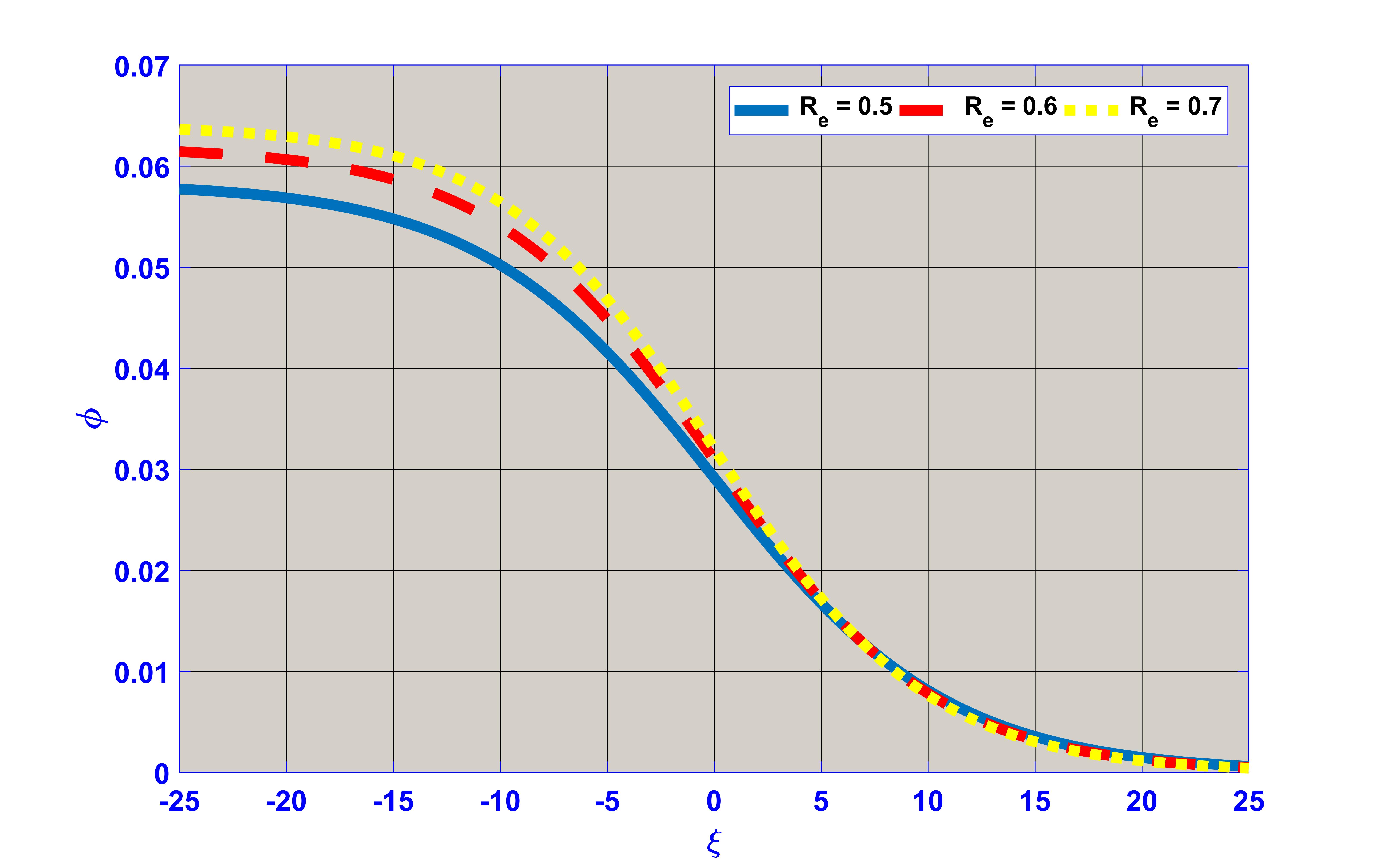}
  \caption{Double layer structure in $(\phi-\xi)$ space for different values of $R_e$ when M = 1.4, $H$=2 and $\mu$=0.6}\label{shockRe.jpg}
\end{figure}

In Figures \ref{solitaryM.jpg} and \ref{solitaryRe.jpg}, the solitons structures are shown. Whereas the Mach number increases the nonlinearity and the soliton amplitude, and the Relativistic degeneracy parameter ($R_e$) increases the dispersive effects. However, Mach numbers have a positive effect in enhancing the double layer separation, but it converges to the asymptotic value in the progressive side. This implying that at higher speed, all particles more together, thus reducing potential separation (fig. \ref{shockM.jpg}). Similar but less pronounced effects are shown for the relativistic parameter ($R_e$) (\ref{shockRe.jpg}). On the contrary reverse behavior is shown for the quantum diffraction parameter (H) (fig. \ref{shockH.jpg}). These results can be related to nature related to the nature of the pseudo-potential.

\section{Conclusion:}
To sum up, we have investigated the properties of small amplitude solitary structure and double layers in an electron-ion degenerate plasma by employing Sagdeev's pseudo-potential approach. From the above investigation, we can say that remarkable potential drops could increase the double layer. The cross-section of the current channel will increase in plasma, incorporating with Goswami et al. \cite{goswami2020solitary}. Furthermore, under any circumstances, the double layer and the transition from sub-thermal to super-thermal electron distribution must coincide if the double layer is strong \cite{smith1978modulation}. This work can shade some light on Buneman instability.
\section{Acknowledgement:}
Authors would like to thank Physics departments of Jadavpur University and Government General Degree College at Kushmandi for providing facilities to carry this work.

\bibliographystyle{IEEEtran}
\bibliography{chap011}
\begin{IEEEbiography}[{\includegraphics[width=1in,height=1.0in,clip,keepaspectratio]{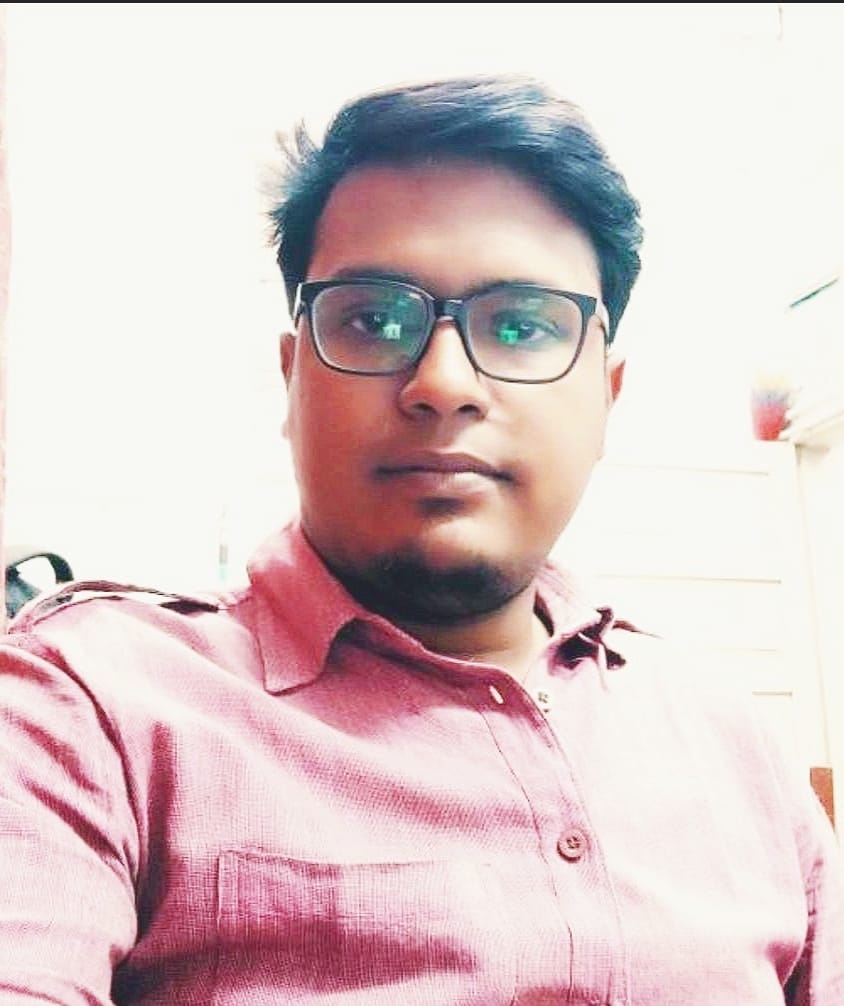}}]
{Jit Sarkar} has completed his Graduation from Department of Physics, Ramakrishna Mission Residential College, Narendrapur. Post-graduation degree was conferred on him in 2015. Currently he is working as a research fellow under Prof B. Ghosh and Dr S. Chandra at Jadavpur University. His field of work based on Astrophysical plasma and Quantum plasma.
\end{IEEEbiography}

\begin{IEEEbiography}[{\includegraphics[width=1in,height=1.0in,clip,keepaspectratio]{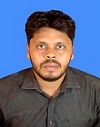}}]
{Jyotirmoy Goswami} is pursuing his Ph.D. from Jadavpur University under the guidance of Prof. Basedev Ghosh and Dr. Swarniv Chandra. He has attended numerous workshops, national and international conferences. He is the winner of the best poster award in the 106th Indian Science Congress.
\end{IEEEbiography}
\vfill
\end{document}